\documentstyle[aps,prl]{revtex}
\input{epsf}

\begin{document}
\draft
\twocolumn
\title{
Relaxation Spectra at the Glass Transition: Origin of Power Laws}
\draft
\author{Michael Ignatiev, Lei Gu  and Bulbul Chakraborty}
\address{
The Martin Fisher School of Physics\\
Brandeis University\\
Waltham, MA 02254, USA}
\date{\today}
\maketitle

\begin{abstract}
We propose a simple dynamical model of the glass transition based on the
results from a non-randomly frustrated spin model which 
is known to form a glassy state below a characteristic quench temperature.
The model is characterized by a multi-valleyed free-energy surface which is
modulated by an overall curvature. The transition associated 
with the vanishing of this overall curvature is reminiscent of the glass
transition.  In particular, the
frequency-dependent response evolves from a Debye relaxation peak to a function
whose high-frequency behavior is characterized by a non-trivial power law.
We present both an analytical form for the response function and numerical
results from Langevin simulations.
 
\end{abstract}

\narrowtext

In supercooled liquids, the glass transition is heralded by anomalously slow 
relaxations\cite{REVIEW}.    The phase below the glass transition temperature is 
non-ergodic 
and the dynamics is characterized by ``aging''\cite{REVIEW}.
Recent experiments indicate that the the approach to the glass transition has some
universal features\cite{NAGEL} when viewed in terms of the frequency-dependent
response of the system. Some of these features are also shared by spin glass
transitions\cite{NAGELSG}. Theoretical research has focussed on spin-glass-like
phenomena in nonrandom
systems\cite{PARISI} and some of these models have been shown to be equivalent
to mean-field
spin-glass  models\cite{FRANZ,PRE95}.  The dynamics of certain frustrated $XY$
models have also been shown to exhibit glassy behavior\cite{KIMLEE}.

In this paper, we present a dynamical model of the glass transition and analyze
the trends in its frequency dependent response.
The model is characterized
by a multi-valleyed 
free-energy surface modulated by an overall
curvature and the glass transition is associated with the vanishing of this
overall curvature.
A multi-valleyed free-energy surface has long been associated with
glasses\cite{REVIEW}. The features
of the present model were deduced from simulations of a
nonrandomly frustrated spin model\cite{MRSLEI}. The frequency-dependent response
of the system
can be calculated in closed form, and its low temperature
approximation becomes exact in the high-frequency
limit.  The results show that, as the overall curvature approaches zero, the 
frequency-dependent susceptibility changes  
from a characteristic Debye relaxation peak to a power-law spectrum reflecting the distribution of
curvatures of the valleys.
The predicted
changes
agree qualitatively with the generic features identified in
experiments\cite{NAGEL}.

We begin with a description of the results of simulations of the frustrated spin system. 
The model considered is an 
Ising antiferromagnet on a deformable triangular lattice.  Detailed
analysis of this model has shown that
there is a first-order transition from the disordered paramagnetic state to
an ordered ``striped'' phase\cite{LEIGU,CHEN}.  
Frustration plays a crucial role
in this phase transition\cite{CHEN}.  The model is characterized by two
order parameters: a staggered magnetization with Ising symmetry and a shear
distortion with three-state Potts symmetry.  

\begin{figure}[thbp]
\epsfxsize=3.0in \epsfysize=2.5in
\epsfbox{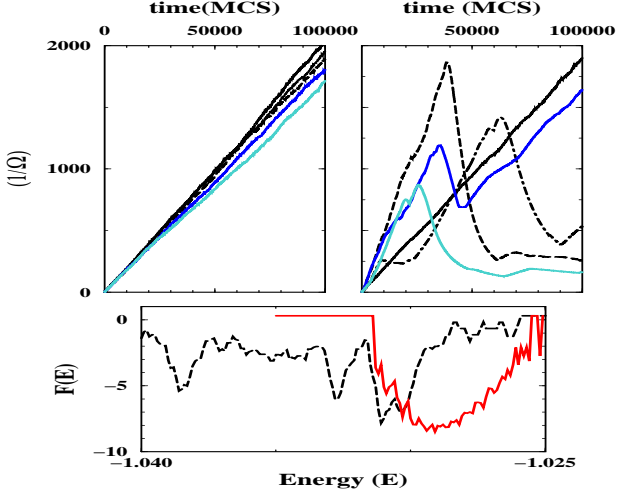}
\caption{Monte Carlo results for ${\Omega}^{-1} (t)$;(a)$T>T^*$ and
(b)$T<T^*$. Fig. 1(c) depicts the structure of the free-energy surface for
$T>T^*$ (solid line) and $T<T^*$ (dashed line).  These results are from a system
of size 96x96}
\label{fig1}
\end{figure}

Monte Carlo simulations\cite{MRSLEI}
have demonstrated that the time evolution of various physical quantities, such as the
average energy, 
undergoes a qualitative change
as the system is quenched below  a characteristic temperature $T^*$, below the
ordering transition\cite{MRSLEI}.
To investigate the
nature of the transition at $T^*$, we computed the fluctuation metric,
$\Omega(t)$\cite{MOUNT}.   This function has been shown to be sensitive to
ergodicity breaking and has been used to study the glassy phase in Lennard-Jones
systems\cite{MOUNT}.   The metric is defined by:
\begin{equation}
\Omega (t)  = <(\nu (t) - \bar{ \nu} )^2>~;~
\nu (t) = {{1} \over {t}} {\int}_{0}^{t} dt^\prime E(t^{\prime}) 
\label{METRIC}
\end{equation}
Here $\bar{\nu}$ is the ensemble average of the energy.
If a system is ergodic, the fluctuation metric decays to zero at times longer than the
equilibration time and behaves as $1/t$ for exponential relaxations\cite{MOUNT}.
Breaking of ergodicity is
signaled by 
the appearance of a non-zero
long-time limit and a trajectory
dependence\cite{MOUNT}.   Fluctuation metrics obtained from our
simulations (Fig. (\ref{fig1})) show that the supercooled state at temperatures
above $T^*$ is 
ergodic but the
state below $T^*$ exhibits broken ergodicity. 
%

A free-energy function 
was obtained from the 
probability distribution of energy\cite{MRSLEI} assuming  
that quasi-equilibrium is reached in each of the valleys\cite{PARISI}.
The resulting function is a one-dimensional projection of the actual,
multidimensional free-energy surface.  As seen in Fig. (\ref{fig1}c), 
at temperatures above $T^*$, there is a well defined valley centered at an
energy corresponding to zero
shear\cite{MRSLEI}.  The curvature of this valley decreases and
subvalley structures become more pronounced as $T^*$ is approached 
and, at the ergodicity-breaking transition, one is left with  a 
multi-valleyed free energy landscape.  These observations are consistent with the idea of a shear 
instability, which is known to exist in this model\cite{CHEN}, occurring at
$T^*$.  The temperature $T^*$ separates  a high-temperature
regime where the valleys with finite shear distortions are not accessible from a
low-temperature regime where these valleys become accessible. The loss of
ergodicity is associated with divergence of the average trapping time in these
valleys and is
reminiscent of weak-ergodicity breaking\cite{MRSLEI,BOUCH}.

Based on these observations of the non-randomly frustrated spin model, we have
constructed a simple dynamical model for the glass 
transition.  The dynamics is described by a Langevin equation based on a
multi-valleyed Hamiltonian $H({\phi})$ where $\phi$ is an order parameter,
similar to the shear distortion in our spin model.
A particularly simple form which 
has the characteristics indicated in our simulations ({\it cf}
Fig. (\ref{fig1}c)), consists of piecewise parabolic 
valleys within a megavalley characterized by an overall curvature.  
The distribution of 
curvatures and depths  of the 
subvalleys, together with the   
overall curvature characterize the model completely.  The glass-transition is
associated with the vanishing of the overall curvature. 


The Hamiltonian describing the piecewise parabolic model is:
\begin{eqnarray}
H[\phi] &=&
\int dt
\sum_{n=0}^{n_{max}}
\Bigl\{{R^2\over 2} \phi^2 +
{r_n\over 2} \bigl(\phi(t)-\phi_n^0\bigr)^2 + c_n
\Bigr\}
\mu_n \nonumber \\
&=& \sum_{n=0}^{n_{max}}
H_n\mu_n,
\end{eqnarray}
where $\{\mu_n\}$ is  the set of characteristic functions on the
non-overlapping segments
$[\phi_n^0-\Delta_n;\phi_n^0+\Delta_n]$, 
covering the domain of $\phi(t)$, 
and $c_n$ are chosen such  that
$H[\phi]$ is continuous.  The parameter $R$ determines the overall curvature and
the individual well curvatures are determined by $r_n$.
The relaxational dynamics of the model is described by the Langevin equation
\begin{equation} 
\label{eq:Langevin}
\dot \phi(t) = 
-
{
	\delta H[\phi]
\over 
	\delta\phi(t)
}+\eta (t),
\end{equation}
with the Gaussian noise
\begin{equation} 
<\eta (t) \eta (t')>=\Gamma\delta (t-t').
\end{equation}
We are interested in studying the frequency dependent response of this system as
the overall curvature is tuned to zero.  The response can be calculated
by constructing an appropriate generating function.
Integrating over the Gaussian noise, leads to the dynamical action\cite{DYNAMIC} 
\begin{equation}
S[\phi] = 
\int dt
\Bigl|
{\partial\over\partial t}\phi(t) + 
{
	\delta H[\phi]
\over 
	\delta\phi(t)
}
\Bigr|^2 ~,
\end{equation}
which defines the generating function
\begin{eqnarray}
Z[h] &=&{\int}{\cal D}\phi (t)
\Bigl\|{\partial\over\partial t}+
{
	\delta^2 H[\phi]
\over 
	\delta\phi^2(t)
}\Bigr\|\nonumber \\
	& &{\exp}\Biggl\{
		-{1\over 2\Gamma}S[\phi] + \int dt h(t)\phi(t)
	\Biggr\}~ .
\end{eqnarray}
For an ergodic system,  the response  can be related to a correlation
function through the fluctuation dissipation theorem.  Approaching the glass
transition temperature from above, we can use this relationship to calculate the
imaginary part of the 
frequency-dependent susceptibility $X (\omega)$ from a knowledge of the
correlation function $\chi$\cite{foot_fluc}:
$$
\chi ={\delta\over\delta h'}<\phi[h]>~;<\phi[h]> = {\delta\over\delta h}\ln Z[h].
$$
Introducing the ``centered'' order parameter,
$x_n=\phi-\phi_n^0$ and ``effective'' subwell curvature $R_n=r_n+R$, and using
the non-overlapping property: $\mu_m\mu_n=\delta_{mn}\mu_n$,  the dynamic
action 
can be rewritten as:
\begin{equation}
S[\phi] = 
\int dt
	\sum_{n=0}^{n_{max}}
	\Biggl|
		\Bigl(
		{\partial\over\partial t} + R_n
		\Bigr)
	x_n(t) + R\phi_0
	\Biggr|^2\mu_n,
\end{equation}
The non-overlapping property also leads to  the generating function being
written as a sum of functions belonging to each
segment: 
$Z[h] = \sum_{n=0}^{n_{max}}Z_n[h]$,
where the subwell generating functions are given by:
\begin{eqnarray}
\label{gen_function}
Z_n[h] &=&
{\int}{\cal D}x_n\mu_n\|\omega_n\|
\exp\Biggl\{
		-{1\over 2\Gamma}\int dt
\Bigl(\omega_n x_n-\Gamma \omega_n^{-1}H_n\Bigr)^2 
\Biggr\}\nonumber \\
& &{\exp}\Biggl(\int dt
	Q_n + {\Gamma (\omega_n^{-1}H_n)^2\over 2}
\Biggr) \nonumber \\
&\equiv&
Z_n^0[h]
\exp\Biggl(\int dt
	Q_n + {\Gamma (\omega_n^{-1}H_n)^2\over 2}
\Biggr).
\end{eqnarray} 
In the above equation, we have introduced the variables:
\begin{eqnarray}
H_n &=& h - {RR_n\phi_n^0\over 2\Gamma} ~,\nonumber \\
\omega_n^2 &=& {\partial^2\over\partial t^2} + R_n^2 ~,\nonumber \\
Q_n &=& h\phi_0^n - {(R\phi_n^0)^2\over 2\Gamma} ~.\nonumber
\end{eqnarray}
If the subwells were not constrained to exist in finite regions of the order
parameter space, then $Z_n^0[h]$ would be a constant and the response of the
system obtained from the resulting generating function would be a sum of Lorentzians.
It will be shown below that this is very nearly the case in the limit of large
frequency  (scale set by the
noise strength $\Gamma$).

Because of the non-overlapping property ({\it cf} Eq. (\ref{gen_function})), the
susceptibility can be expressed in terms $\chi_n$ and $\bar\phi_n$ which
denote  the 
susceptibility and average order parameter of the individual, decoupled wells.
The calculation of ${\chi}_n$ defies any simple analytic approach, because
of the finite region of functional integration in the partition function.
However, if the noise is small enough and the system never escapes its
subwell, we can neglect all these effects of finiteness of the wells.
It can be shown\cite{foot_exp} that the error vanishes as we increase 
$\xi_n=\omega_n^2\Delta_n^2 / 2\Gamma$. In the zero noise limit all
$Z_n^0[h]$ are the same and do not depend on $h$. For finite noise, the same
effect can be achieved by increasing the frequency associated with the response
function and therefore, the low-noise approximation is less restrictive for high frequencies.

The simplest case is when the overall curvature, $R$ vanishes. In this limit,
which we associate with the glass transition,  the frequency-dependent response
$\chi_n$ is given by:
\begin{equation}
\label{eq:chin}
\chi_n (\omega)= {\Gamma\over\omega_n^2} ~ .
\end{equation}
Here, ${\omega}_n ^2 = {\omega}^2 + {R_n}^2$.
The ``effective'' susceptibility $\chi=<\chi_n>$ is given by the weighted sum
over all the subwells:
$$
\chi = \sum_n p_n \chi_n,\ \ \ p_n = {Z_n\over Z}.
$$ 
Since 
$\chi_n$ depends on $n$ only through the curvature $r_n$ ($R=0$), the
susceptibility 
$\chi$ can be expressed in terms of the probability of encountering a given curvature:
\begin{equation}
\label{eq:Zero_curv_chi}
\chi = 
\int{dr} {P(r)\over\omega^2+r^2}~ ; P(r) \equiv {\sum}_n p_n \delta (r-r_n)
\end{equation}
In general, $P(r)$ depends on the distribution of curvatures and the `time'
spent in a well with given curvature.  The latter is determined by the depth of the
well and the overall curvature. However, in the low-noise regime, where we have
argued that $p_n$ is independent of $n$,
$P(r)$ reflects the
`intrinsic' distribution of curvatures in the model Hamiltonian.  If this
distribution is described by a power law: $r^{\alpha}$, then within our
approximation which is exact in the high-frequency limit, the response of the
system reflects this power law.
\begin{equation}
\label{power_law}
\chi(\omega) = \int{dr} {r^{\alpha}\over\omega^2+r^2} \simeq {\omega}^{\alpha -
1}
\end{equation}
 
When the curvature $R$ is non-zero, the probabilities, $p_n$, 
get multiplied by $\exp(-R^2{\phi_n^0}^2/2\Gamma)$ and 
even for extremely small overall curvature, $R$, the distribution $P(r)$ changes
considerably due to a different sampling of the wells.  As $R$ increases only
subwells centered close to $\phi = 0$ are sampled appreciably and, in the limit
of large $R$,  the response
of the system is dominated by a single parabolic well (Debye response).
The intrinsic
distribution of curvatures is reflected in the susceptibility only in the limit
of $R = 0$.
This result shows that by tuning the overall curvature of a
multi-valleyed free-energy surface the high-frequency response of the system can
be changed from a pure Debye spectrum ($\chi \simeq {\omega}^{-2}$) to a
nontrivial power law which reflects the distribution of curvatures of the subwells. 
\begin{figure}[tbhp]
\epsfxsize=3.0in \epsfysize=2.5in
\epsfbox{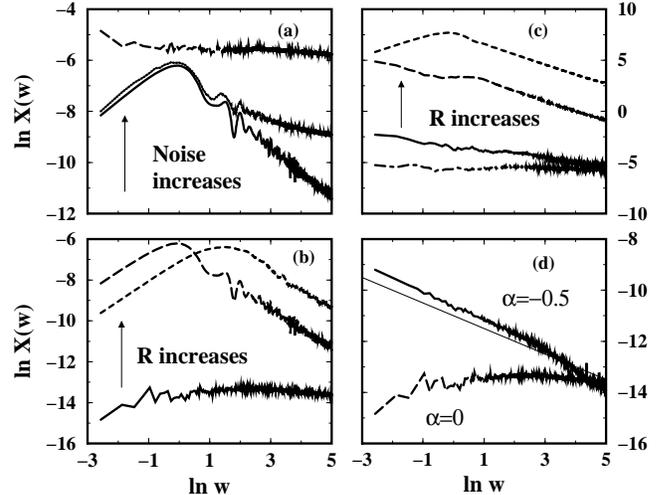}
\caption{Plots of $X(\omega)$. (a) $R = 0.001$,
increasing values of $\Gamma$; 0.01(solid), 0.1(dotted), 0.5(dashed); 
(b) $\Gamma = 0.01$ increasing values of $R$; 0(solid), 0.001(dashed),
0.01(long-dashed);(c) $\Gamma = 0.5$, at different values of $R$;
0.001(dot-dashed), 0.05(solid), 0.1(long-dashed), 0.2(dashed); and 
(d) Susceptibility  at $R=0$ and $\Gamma = 0.01$ for two different values of
$\alpha$. The straight line has a slope of $-0.5$.}
\label{fig2}
\end{figure}

To demonstrate these features explicitly  and to  analyze the effect of large noise,  we
performed simulations of the Langevin equation
(\ref{eq:Langevin}). The subwells were chosen to be distributed uniformly
($\alpha = 0$) and the
largest subwell curvature was determined by  considering the numerical 
stability of the system which required that in each well $\dot\phi\Delta
t < 2\Delta_n$
In each run the starting point was near the bottom of a
randomly chosen subwell. We calculated the susceptibility as an average
over a large number ($\sim 1000$) of runs and each run was $\sim 10000$
Monte Carlo Steps long. The results are shown in the Fig (\ref{fig2}).

For zero overall curvature, irrespective of the noise strength, the
susceptibility reflects
the distribution of the subwell curvatures (Fig. (\ref{fig2})d). 

For a non-zero overall curvature, one can distinguish between two distinct regimes
(a) low noise such that there is no escape from wells and (b) noise large enough
to make escape possible
The behavior in the (a) regime is summarized in Figs. (\ref{fig2}a) and
(\ref{fig2}b).
At these noise strengths, a finite value of the overall
curvature leads to a large damping of the contributions  
from all subwells centered away from $\phi = 0$, 
and one observes a
superposition of very few Lorentzian peaks (Fig. (\ref{fig2}b)).  
If the noise strength is increased, at small $R$, the  susceptibility starts to  pick up
contributions from all the subwells and the response becomes non-Debye like
({\it cf} Fig. (\ref{fig2}a)). 

The response of the system is qualitatively different in regime (b), as
illustrated in Fig (\ref{fig2}c).
The  finiteness of the
wells becomes a crucial
factor and, at some minimum value of $R$, escape from the 
most distant wells became possible.  
The dominant contribution to the 
susceptibility arises from the relaxation towards $\phi = 0$ and  at large $R$
there is a crossover to a  single Debye peak with no evidence of the internal structure.
In the intermediate regime, the susceptibility reflects a superposition
of both probing the distribution and relaxation towards the center.
In this case the shape of the curve is sensitive to the details
of all the distributions: centers of subwells, their widths
(depths) and curvatures.

The numerical and analytical results demonstrate that by tuning $R$ or the noise
strength $\Gamma$, one can reach a special ``glassy'' phase with a frequency-dependent
susceptibility obeying a non-trivial power law (Eq. (\ref{power_law}).
For large noise and curvature, one observes a Debye peak determined by the
overall curvature $R$. Starting from this liquid-like
phase,  the power-law regime is reached  by decreasing the curvature.   In the low-noise
limit (regime (a)),  the phase with the Debye-like response resembles 
a frozen amorphous or polycrystalline state since there is no escape from
subwells.  The location of the Debye peak is determined not by the overall
curvature $R$ but by the average curvature of the subwells located near $\phi =
0$.  This frozen phase is unusual since the system still relaxes in a multitude
of subwells and is far from equilibrium.  The frozen phase can be made to
approach the ``glassy'' phase by increasing the noise strength.  The power-law
behavior characterizes the glass and is not observed in either the liquid or the
frozen phase.

The transition
from the ``liquid'' to the ``glassy'' phase, as depicted in Fig. (\ref{fig2}c)
is similar to experimental observations in supercooled
liquids\cite{NAGEL,NAGELSG}.   In both theory and experiment,  the shift of the
Debye peak towards lower frequencies is accompanied by a change in the power law
which characterizes the high-frequency response.  The scaling features observed
in experiments\cite{NAGEL} are intriguing and probably reflect some interesting
physical features of glass formers.  This physics would dictate the type of
distributions of curvatures and heights which enter our theoretical model.  As
mentioned before, all these distributions are crucial in determining the
behavior of $\chi (\omega)$ in the regime intervening between the large $R$ and
$R=0$ phases.  We have made no attempt to tailor these distributions to
reproduce the scaling behavior observed in experiments. 

In conclusion, we have shown that our model of the glass transition predicts
changes in the high-frequency response of a system approaching the glass
transition.   This model can be taken only as a schematic picture of a real
glass transition but suggests that a multi-valleyed free-energy surface with an
overall curvature can be taken as a useful template for glassy systems.  The
transition associated with the vanishing of the curvature is a new type of
critical point and our
analysis shows that this ``phase transition'' is characterized by a non-trivial, high-frequency
response.


The authors wish to thank R. Stinchcombe and N. Gross for helpful
conversations. The work of M.I. and B.C.  was supported 
in part by the NSF grant DMR-9520923.  The work of Lei Gu was supported in part
by the DOE grant DE-FG02-ER45495.


\begin{references}
\bibitem{REVIEW}M. D. Ediger, C. A. Angell and Sidney R. Nagell, 
J. Phys. Chem {\bf 100}, 13200 (1996); J. J\"{a}ckle, Rep. Prog. Phys. {\bf
49}, 171 (1986) and references therein.
\bibitem{NAGEL} P. K. Dixon {\it et al}, Phys. Rev. Lett {\bf 65}, 1108
(1990); N. Menon and S. R. Nagel,  Phys. Rev. Lett {\bf 74}, 1230 (1995).
\bibitem{NAGELSG}D. Bitko {\it et al}, Europhys. Lett. {\bf 33}, 489 (1996).
\bibitem{PARISI} J. P. Bouchaud and M. M\'{e}zard, J. Phys. I (France) {\bf 4},
1109 (1994); G.Parisi, Report no. cond-mat/9412034
\bibitem{FRANZ} S. Franz and J. Hertz, Phys. Rev.Lett. {\bf 71}, 2114 (1995).
\bibitem{PRE95}J. P. Bouchaud, L. Cugliandolo, J. Kuchran and M. M\'{e}zard,
Report no. cond-mat/9511042.
\bibitem{KIMLEE}B. Kim and S. J. Lee, Phys. Rev.Lett. {\bf 78}, 3709 (1997).
\bibitem{MRSLEI}Lei Gu and Bulbul Chakraborty, Report no. cond-mat/9612103,{\it
Proceedings of MRS Symposium on Glasses}, eds. K. L. Ngai and C. A. Angell, 1996.
\bibitem{LEIGU} Lei Gu {\it et al}, Phys. Rev. B {\bf 53}, 11985 (1996)
\bibitem{CHEN} Z. Y. Chen and Mehran Kardar, J. Phys. C: Solid State Phys. {\bf
19}, 6825 (1986).
\bibitem{MOUNT}D. Thirumalai and R. D. Mountain, Phys. Rev. A {\bf 42}, 4574 (1990).
\bibitem{BOUCH} J. P. Bouchaud, J. Phys. I (France) {\bf 2}, 1705 (1992).
\bibitem{DYNAMIC} J. Zinn-Justin, {\it Quantum Field Theory and Critical
Phenomena} (Clarendon Press, 1989) Chapter 3.
\bibitem{foot_fluc}In the following, we have used the word susceptibility
interchangeably with the correlation function.  In frequency space, the
difference between them is in a factor $\omega$.
\bibitem{foot_exp}M. Ignatiev, unpublished
\end{references}
\end{document}